\documentclass[
    twoside,
    aps,
    prd,
    superscriptaddress,
    twocolumn,
    a4paper,
    floatfix,
    longbibliography,
    nofootinbib
]{revtex4-2}

\usepackage[american]{babel}
\usepackage[utf8x]{inputenc}

\usepackage{grffile} 

\usepackage{newtxtext,newtxmath}
\usepackage{microtype}

\usepackage{amsmath}
\usepackage{bbm}
\usepackage{bm}
\usepackage{mathtools}
\usepackage{dsfont}
\usepackage{braket}
\usepackage{cancel}
\usepackage{slashed}

\usepackage{graphicx}
\usepackage[svgnames]{xcolor}

\usepackage{siunitx} 

\usepackage{ragged2e}
\usepackage{array}
\usepackage{tabularx}
\usepackage{booktabs}

\definecolor{cset-aps-blueberry}{RGB}{28,128,158}
\definecolor{cset-aps-blue}{RGB}{46,44,184}
\definecolor{cset-aps-turquoise}{RGB}{0,67,88}
\definecolor{cset-aps-limegreen}{RGB}{190,219,67}
\definecolor{cset-aps-green}{RGB}{31,138,112}
\definecolor{cset-aps-yellow}{RGB}{255,225,25}
\definecolor{cset-aps-orange}{RGB}{253,116,0}
\definecolor{cset-aps-red}{RGB}{219,0,43}

\definecolor{cset-aps-kobalt-medium}{RGB}{62,54,222}
\definecolor{cset-aps-kobalt-dark}{RGB}{28,24,150}

\usepackage{tikz}

\usepackage{pgfplots}
\pgfplotsset{%
    every axis legend/.append style={%
        cells={anchor=west},
        at={(0.96,0.04)},
        anchor=south east,
        font=\scriptsize,
        },
    every axis/.append style={%
        yticklabel style={%
            /pgf/number format/fixed zerofill,
            /pgf/number format/precision=2},
        },
    width= \textwidth,
    height=8cm,
    xmajorgrids=true,
    xminorgrids=false,
    minor x tick num=1,
}
\usepgfplotslibrary{external}

\usetikzlibrary{decorations}

\usepackage{pict2e,picture}

\makeatletter
\DeclareRobustCommand{\Arrow}[1][]{%
\check@mathfonts
\if\relax\detokenize{#1}\relax
\settowidth{\dimen@}{$\m@th\rightarrow$}%
\else
\setlength{\dimen@}{#1}%
\fi
\sbox\z@{\usefont{U}{lasy}{m}{n}\symbol{41}}%
\begin{picture}(\dimen@,\ht\z@)
\roundcap
\put(\dimexpr\dimen@-.7\wd\z@,0){\usebox\z@}
\put(0,\fontdimen22\textfont2){\line(1,0){\dimen@}}
\end{picture}%
}
\makeatother

\usepackage{hyperref}
\hypersetup{%
    colorlinks=true,
    linkcolor={cset-aps-red},
    linkbordercolor={cset-aps-red},
    filecolor={cset-aps-orange},
    filebordercolor={cset-aps-orange},
    citecolor={cset-aps-kobalt-medium},
    citebordercolor={cset-aps-kobalt-medium},
    urlcolor={cset-aps-kobalt-dark},
    urlbordercolor={cset-aps-kobalt-dark},
    menucolor={cset-aps-limegreen},
    menubordercolor={cset-aps-limegreen},
    breaklinks=true,
    pdfborderstyle={/S/U/W 2},
    pdfpagemode=UseOutlines,
    pdfstartpage={1},
}

\newcommand{\ee}{\text{e}}
\newcommand{\ii}{\text{i}}
\newcommand{\dd}{\text{d}}
\newcommand{\vect}[1]{\boldsymbol{#1}}

\usepackage{wasysym}

\usepackage[clock]{ifsym}
\usepackage{lipsum}

\newcommand{\orcid}[1]{\href{https://orcid.org/#1}{\includegraphics[width=7pt]{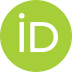}}}

\begin{document}

\title{Universality-of-clock-rates test using atom interferometry with $T^{3}$ scaling}
\collaboration{This article has been published in \href{https://doi.org/10.1103/PhysRevD.107.064007}{Physical Review D \textbf{107}, 064007 [2023]}}

\newcommand{\affHAN}{\address{Institut f{\"u}r Quantenoptik, Leibniz Universit{\"a}t Hannover, Welfengarten 1, D-30167 Hannover, Germany}}
\newcommand{\affULM}{\address{Institut f{\"u}r Quantenphysik and Center for Integrated Quantum Science and Technology (IQ\textsuperscript{ST}), Universit{\"a}t Ulm, Albert-Einstein-Allee 11, D-89069 Ulm, Germany}}
\newcommand{\affTUDa}{\address{Technische Universit{\"a}t Darmstadt, Fachbereich Physik, Institut f{\"u}r Angewandte Physik, Schlossgartenstr. 7, D-64289 Darmstadt, Germany}}
\newcommand{\affDLR}{\address{Institute of Quantum Technologies, German Aerospace Center (DLR), S\"{o}flinger Stra\ss e 100, D-89077 Ulm, Germany}}
\newcommand{\affHagler}{\address{Hagler Institute for Advanced Study and Department of Physics and Astronomy, Institute for Quantum Science and Engineering (IQSE), Texas A{\&}M University, College Station, Texas 77843-4242, USA}}
\newcommand{\affTexas}{\address{Texas A{\&}M AgriLife Research, Texas A{\&}M University, College Station, Texas 77843-4242, USA}}

\author{Fabio Di Pumpo\,\orcid{0000-0002-6304-6183}}
\email{fabio.di-pumpo@uni-ulm.de}
\email{fabio.di-pumpo@gmx.de}
\author{Alexander Friedrich\,\orcid{0000-0003-0588-1989}}
\author{Christian Ufrecht\,\orcid{0000-0003-4314-9609}}
\affULM
\author{Enno Giese\,\orcid{0000-0002-1126-6352}\,}
\affTUDa
\affHAN

\begin{abstract}
\noindent
Metric descriptions of gravitation, among them general relativity as today's established theory, are founded on assumptions summarized by the Einstein equivalence principle (EEP).
Its violation would hint at unknown physics and could be a leverage for the development of quantum gravity.
Atomic clocks are excellent systems to probe aspects of EEP connected to (proper) time and have evolved into a working horse for tests of local position invariance (LPI).
Even though the operational definition of time requires localized and idealized clocks, quantum systems like atoms allow for spatial superpositions that are inherently delocalized.
While quantum experiments have tested other aspects of EEP, no competitive test of LPI has been performed or proposed allowing for an intrinsic delocalization.
We extend the concepts for tests of the universality of clock rates (one facet of LPI) to atom interferometry generating delocalized quantum clocks.
The proposed test depends on proper time with a favorable scaling and is, in contrast to fountain clocks, robust against initial conditions and recoil effects.
It enables optical frequencies so that the projected sensitivity exceeds the one of state-of-the-art localized clocks.
These results extend our notion of time, detached from classical and localized philosophies.
\end{abstract}

\maketitle
\section{Introduction}
\begin{figure*}
    \centering
    \includegraphics{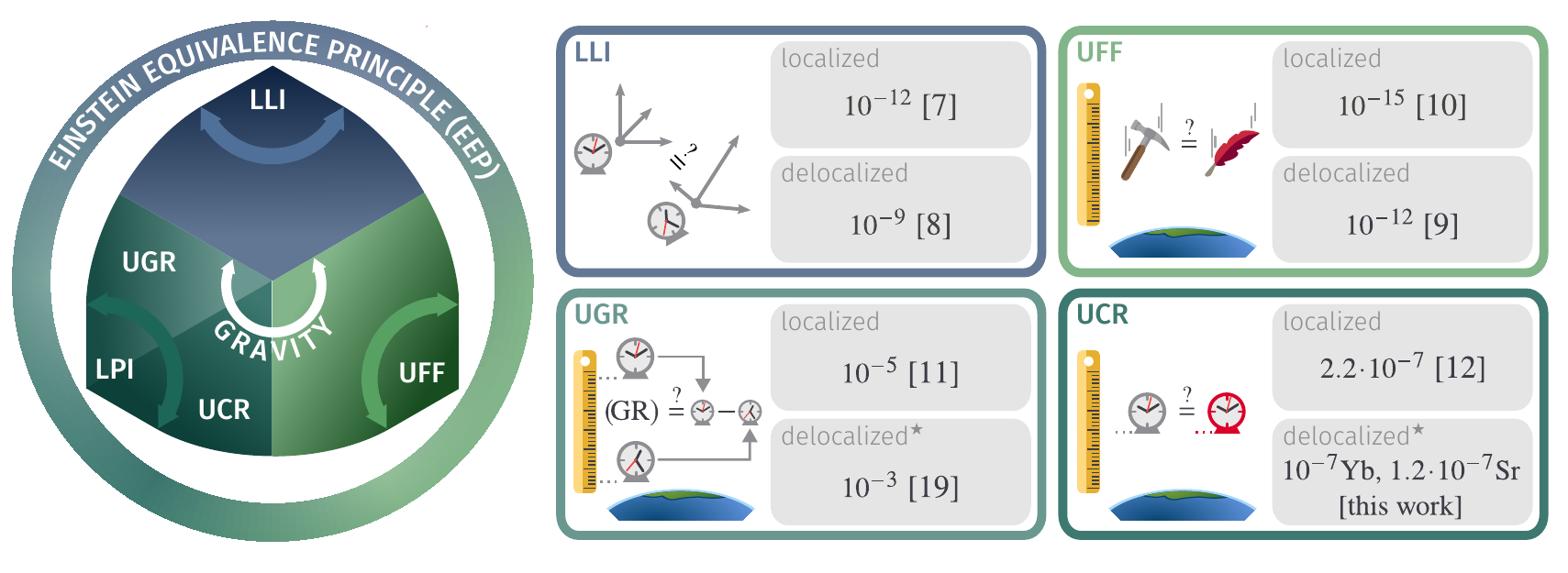}
    \caption{
    Key fundamental assumptions that constitute the Einstein equivalence principle (EEP).
    It rests on three pillars: 
    local Lorentz invariance (LLI), the universality of free fall (UFF), and local position invariance (LPI).
    Whereas LPI and UFF are inherently connected to gravity, LLI only entails special relativity as a special case of GR.
    The right part visualizes prominent experiments together with bounds set by localized and delocalized schemes, where asterisks indicate projected values.
    The corresponding violation parameters can be found in the references given in brackets.
    Tests of LLI search for a dependence on the choice of temporal and spatial coordinates, probing for preferred reference frames as implied by deformed coordinate systems.
    The provided bound value is given by lunar laser ranging, whereas the less restrictive bound of delocalized tests is based on atom interferometry.
    UFF states that the behavior of test bodies in gravitational fields does not depend on their composition. 
    It can be tested by drop experiments, where the acceleration of different objects is compared.
    Here, the localized experiment gives the bound set by space missions, whereas the delocalized test is based on atom interferometry with different species.
    Tests of LPI search for a dependence on the absolute position (or starting time) of the performed experiment and has two prominent facets: the universality of the gravitational redshift (UGR) and the universality of clock rates (UCR).
    For UGR, this dependence is probed by comparing the time difference of two identical clocks at different heights in a gravitational field to the prediction of GR.
    No delocalized version of such an experiment has been performed, but only a lower sensitivity seems to be feasible with quantum-clock interferometry.
    In contrast, UCR can be tested by comparing the time of two clocks with different compositions on the same height.
    As the only principle, UCR allows for a delocalized test proposed in this work that is based on atom interferometry, where the projected sensitivity exceeds the one of classical or localized experiments.
    }
    \label{fig:EEP}
\end{figure*}

Fundamental physics research strives for ever more precise tests of the Einstein equivalence principle (EEP), a cornerstone of general relativity (GR)~\cite{Einstein1905,Einstein1907,Einstein1911}, to verify the universality of the gravitational coupling to test bodies~\cite{Will2014}, translating into three basic assumptions~\cite{DiCasola2015}:
local Lorentz invariance (LLI), universality of free fall (UFF), and local position invariance (LPI); see Fig.~\ref{fig:EEP} that visualizes these aspects as well as their differences, and highlights current state-of-the-art tests.

Of these aspects, LLI is the basis for special relativity as a special case of GR when sufficiently small spacetime regions are considered.
By that, it is not inherently connected to gravity.
LLI can be tested by searching for preferred reference frames, for example by comparing transition frequencies of atoms (and the associated natural constants) measured in different reference frames~\cite{Hohensee2013}.
Analogous setups to test other constants of nature have been implemented as well~\cite{Bourgoin2016}.
Even atom-interferometric tests of LLI~\cite{Chung2009} relying on an intrinsic delocalization of quantum objects have been performed.
Contrarily, UFF (or weak equivalence principle) is a purely gravitational principle and states that the motion of different test bodies in a gravitational field is the same, independent of their composition.
Quantum tests of UFF based on atom interferometry~\cite{Asenbam2020} make use of the delocalization of the atom and have evolved into a competitive alternative to classical drop experiments~\cite{Touboul2017}.

Whereas both LLI and UFF have been verified to high precision~\cite{Bourgoin2016,Touboul2017}, this statement is true for LPI only to a lesser degree~\cite{Delva2018,Ashby2018}.
LPI states that (local) experiments with test bodies are independent of where (and when) they are performed in a gravitational field.
As a consequence, the results are independent of the test objects' compositions.
It has two prominent facets~\cite{Will2014}: universality of the gravitational redshift (UGR) and universality of clock rates (UCR).
We focus on UCR which states that ticking rates defined by two \emph{different} physical systems moving along the \emph{same worldline} are universal, i.\,e., independent of the composition of the systems and their spatial location.
Contrarily, UGR demands that ticking rates defined by two \emph{identical} systems placed at \emph{different heights} are universal.
For a UGR test, two (local) experiments at different positions have to be synchronized (e.g. by classical communication).
The result has to be compared to a theoretical prediction (usually GR), which requires measurements of distances or similar.
In contrast, UCR is a null test where no classical communication and measurements of distances are necessary.
Since different species at the same location are compared, such tests possibly allow for larger effects and may \emph{a priori} test other theories than UGR tests.
Both principles are based on different couplings of internal energies to gravity~\cite{Sinha2011,Zych2011,[][ and references therein.]Pikovski2017,Loriani2019}.
Although traditionally associated with localized\footnote{
In the context of GR and EEP, the term \emph{local} refers to sufficiently small spacetime regions connected to a local (proper) reference frame.
In contrast, a \emph{localized} object is centered around a single spacetime point.
These two terms are not equivalent, since also experiments with delocalized quantum objects can be performed locally (in a local reference frame).
} systems such as clocks, localization is not a crucial factor for such tests.
Indeed, UGR tests based on delocalized schemes such as quantum-clock interferometry and atom interferometry with internal transitions~\cite{Roura2020,Ufrecht2020,DiPumpo2021} have been proposed but not performed.
However, their estimated sensitivity is not yet competitive to their localized counterparts~\cite{Delva2018}.

\subsection{Overview and main results}
Here, we combine the underlying concepts of both atomic clocks~\cite{Nicholson2015,Brewer2019,Oelker2019,Madjarov2019} and light-pulse atom interferometry~\cite{Kasevich1991,Cronin2009} to propose the first quantum sensor for UCR which allows for delocalized superpositions, extending our understanding of ticking rates of clocks.
Thus, we complement UGR and UFF tests based on such delocalized quantum superpositions to test GR.
Even more, because the UCR test can be made robust against experimental imperfections, its projected sensitivity exceeds the one of state-of-the-art fountain clocks~\cite{Ashby2018} and therefore represents the first EEP test based on atom interferometry that has the potential to outperform its localized counterpart, see Fig.~\ref{fig:EEP}.
\vspace{-1em}
\subsection{Proper time of a freely falling particle}
The geometric formulation of GR implies that the parametrization of a worldline is not unique.
However, a comoving observer has a distinguished notion of time, namely proper time recorded by a comoving idealized clock, which is inherently connected to spacetime.
Thus, proper time can be used to test the universality of the gravitational coupling to test bodies assumed by EEP.
It is determined by the spacetime geometry and varies with the path length along the worldline.
When expressed in terms of laboratory coordinates, proper time in the weak-field and low-velocity limit~\cite{Will2014} reduces to
\begin{equation}
\label{eq:PropTime}
    \tau =\int\limits_0^T \dd t\left(1-\frac{\dot{\vect{r}}^2}{2c^2}+\frac{\vect{g}\vect{r}}{c^2}\right) =  T - \frac{\vect{g}^2}{3c^2} T^3 = T - \Delta \tau
\end{equation}
for a classical particle along the trajectory $\vect{r}(t)=-\vect{g}t^2/2$ in a linear gravitational field with acceleration $\vect{g}$.
Here, $c$ is the speed of light and $t$ is the laboratory time ranging from $0$ to $T$. 
In addition to $T$ we find a contribution $\Delta\tau=T^3 \vect{g}^2/(3c^2)$ that scales cubically with the laboratory time~\cite{Penrose2014,Marletto2020}.

Pointlike two-level atoms are the simplest physical realization of clocks.
However, real atoms exhibit quantized center-of-mass (c.m.) motion, such that even a narrow wave packet probes the vicinity of the worldline. 
Consequently, the evolution of quantum systems encodes information about the initial c.m. wave packet, internal degrees of freedom, and the spacetime geometry into quantum observables, like the phase in interference experiments. 
Observables isolating proper time in a (quantum) experiment are sufficient to introduce the notion of clocks and are susceptible to UCR violations, e.g. if internal states are affected differently by gravity.
Based on UCR violating models~\cite{Alves2000,Damour1999,Damour2010,Damour2012,Roura2020,Ufrecht2020,DiPumpo2021,DiPumpo2022}, modified gravitational accelerations for each internal atomic state are reflected in a violation parameter $\alpha$ as detailed below.

\section{Phases of clocks and proposed interferometer scheme}
As a prime example for UCR tests, we present the phase measured by atomic fountain clocks and identify its dependence on initial conditions.
Moreover, we introduce an atom interferometer and show that its phase also contains UCR-violating contributions, however, being less sensitive to initial conditions.

\subsection{Phase of fountain clocks}
In a Ramsey sequence~\cite{Ramsey1950}, a superposition of two internal states initiated by a $\pi/2$ pulse is read out by another $\pi/2$ pulse after a time interval $T$.
By measuring interference fringes encoded into the population of the excited state, one obtains the phase difference $\varphi_\text{\clock}^{\vphantom{(1)}}$ acquired between both internal states.
The phase of a freely falling clock, e.g. implemented in fountains~\cite{Ashby2018,McGrew2019,Lange2021}, takes the form (see the \hyperref[app:a]{Appendix})
\begin{align}
\label{eq:phase_clock}
\begin{split}
    \frac{\varphi_\text{\clock}^{\vphantom{(1)}}}{ -\Omega T} =&\left(1+\frac{\alpha}{2}\right)\left[\frac{\vect{g}(2\vect{r}_0+\vect{v}_0 T)}{c^2}-\frac{\Delta \tau}{T}\right] - \frac{\vect{g}\vect{r}_0}{ c^2}- \frac{ \langle \vect{v}_0^2\rangle}{2 c^2}
\end{split}
\end{align}
where $\Omega$ is the frequency of the clock transition and the light field is resonant to it\footnote{The preparation of identical clocks on different heights separated by a distance $\vect{r}_0$ with same initial velocity leads to tests of the universality of gravitational redshift~\cite{DiPumpo2021}, which emerges directly from Eq.~\eqref{eq:phase_clock}.}.
Here, recoil effects and finite pulse durations have been neglected.
We observe the phase contribution $\Omega \Delta \tau \sim \Omega T^3 \vect{g}^2/c^2$, and a UCR-violating factor $1+\alpha/2$.
In GR we find $\alpha=0$, so this phase difference reduces to a measurement of proper time including initial conditions $\vect{r}_0$ and $\vect{v}_0$, complemented by wave-packet effects.
The latter arise from different dispersion relations of internal states and enter via the second moment of the initial velocity $\langle \vect{v}^2_0\rangle$, that includes contributions from the second-order Doppler shift~\cite{Daams1974,Brewer2019}. 
The initial conditions may vary between experimental runs of such clock-based tests and consequently limit their accuracy.

\subsection{Proposed UCR-sensitive atom interferometer}
\begin{figure}
    \centering
    \includegraphics{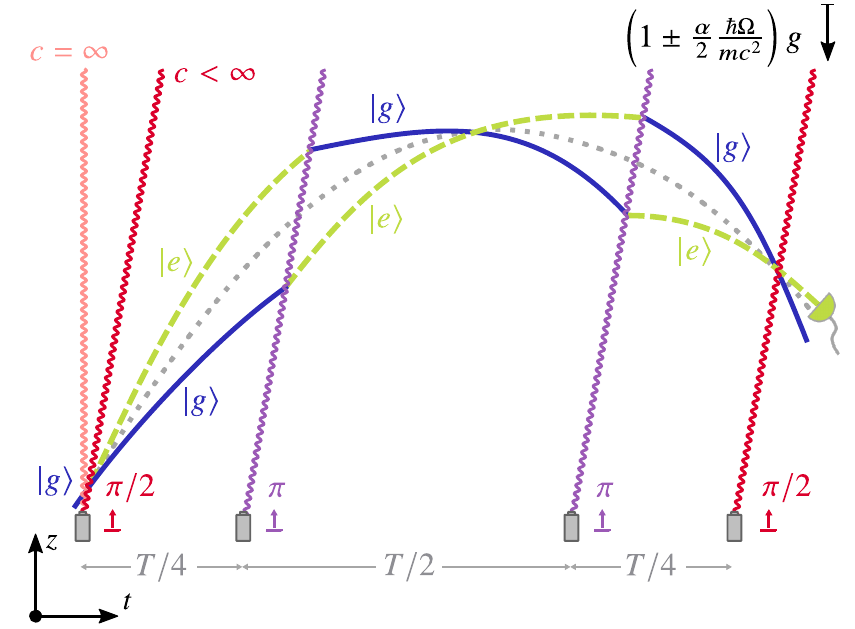}
    \caption{
    Spacetime diagram of the atom interferometer suitable for UCR tests.
    An atom enters in its ground state.
    A first $\pi/2$ pulse (red) generates a superposition of ground state $\ket{g}$ (blue solid line) and excited state $\ket{e}$ (green dashed line) and may impart a momentum difference $\hbar \vect{k}$ between both states (e.g., induced by two-photon Raman or optical single-photon transitions).
    Both internal states are entangled with the c.m. motion of the atom, and consequently it travels in a superposition of two branches.
    They are redirected by two $\pi$ pulses (purple) which invert the momenta and internal states, so that both branches overlap perfectly and interfere at the final $\pi/2$ pulse.
    The finite propagation velocity of the laser pulses is indicated by inclined lines.
    For comparison, the first pulse is also drawn as a straight line representing an infinitely fast pulse.
    The population of the excited state is detected.
    During the experiment, the atom is accelerated by the gravitational acceleration $\vect{g}$, modified by possible violations of the equivalence principle encoded in prefactors $1\pm\alpha\hbar\Omega/(2m c^2)$ with violation parameter $\alpha$, transition frequency $\Omega$, and atomic mass $m$.
    }
    \label{fig:t3-geometry}
\end{figure}

We propose an alternative geometry based on atom interferometry, see Fig.~\ref{fig:t3-geometry}.
Apart from recoilless transitions~\cite{Alden2014}, the $\pi/2$ pulses used in atom interferometers not only generate internal superpositions, but also superpositions of two momenta separated by the (effective) photon recoil $\hbar\vect{k}$~\cite{Giese2015,Kleinert2015} and lead to two separated branches.
This spatial superposition highlights the delocalization and quantum nature of atom interferometers: a single worldline cannot be connected to the atom even for narrow wave packets.
We include two $\pi$ pulses that invert momenta and internal states at times $T/4$ and $3T/4$ so that the geometry closes in phase space at the end of the sequence~\cite{Roura2014}. 
This scheme, known also as butterfly or figure-eight interferometer~\cite{Clauser1988,Marzlin1996,McGuirk2002,Kleinert2015}, is susceptible to quadratic potentials like gravity gradients, while linear accelerations are suppressed.
Neglecting finite pulse durations, laser phase noise, and other parasitic effects, we find the phase (see the \hyperref[app:a]{Appendix})
\begin{align}
\label{eq:PhButterfly}
\begin{split}
    \varphi= \frac{\Omega T^3\vect{g}^2}{16c^2} \left(1+\frac{\alpha}{2}\right)-\frac{3\omega_{\vect{k}} T^3g^2_{\vect{k}}}{32c^2}+\frac{\omega_{\vect{k}}\vect{v}_r\left[\vect{r}_0+\vect{r}_T\right]}{2c^2}. 
 \end{split}
\end{align}
The last two terms arise from the finite speed of light~\cite{Dimopoulos2008,Tan2016,Tan2017}, including the midpoint trajectory $\vect{r}_T=\vect{r}_0+\vect{v}_0 T+\vect{v}_r T/2-\vect{g}T^2/2$ at time $T$ and the gravitational acceleration $g_{\vect{k}}=\vect{k}\vect{g}/\left|\vect{k}\right|$ in  direction of light propagation.
Here, $\vect{v}_r=\hbar \vect{k}/m$ is the recoil velocity, with $m$ being the atom's mean mass between both internal states. For single- and two-photon transitions, we identify $\omega_{\vect{k}}=c\left|\vect{k}\right|$.
Contrarily, using recoilless transitions with $\vect{k}=\vect{v}_r=0=\omega_{\vect{k}}$, all finite-speed-of-light effects vanish.
In this case, the geometry isolates the phase $\Omega \Delta \tau (1+\alpha/2)$ that scales with $T^3$.
It effectively corresponds to a falling clock, not operated in a typical Ramsey sequence, and can be associated with UCR tests.

Interferometer phases scaling with $T^3$ have been proposed~\cite{Zimmermann2017,Zimmermann2019,Zimmermann2021} to detect differential accelerations between both interferometer branches.
While any branch-dependent acceleration displays such a scaling~\cite{McDonald2014}, nonrelativistic state-dependent differential accelerations are isolated in the discussed geometry and are used to determine magnetic field gradients~\cite{Amit2019}.
In this case, the measured phase is not caused by spacetime but by a differential motion between internal states.
In contrast, the gravitational acceleration $\vect{g}$ in GR is independent of the internal state. 
The relativistic $T^3$-phase~\cite{Marletto2020} is solely caused by states with different mass-energy coupling to gravity.
However, differential accelerations, for example induced by magnetic fields, black body radiation, rotations or gravity gradients, may cloak UCR violations~\cite{Lan2012,Dickerson2013,Roura2014,Roura2017,Overstreet2018,Ufrecht2021,Zimmermann2021}, as discussed below.

\section{Tests of UCR}
To establish a common framework for UCR tests, we compare the phases of two nonidentical fountain clocks and derive a consistent UCR violation parameter, which, however, is masked by the dependence on initial conditions.
We identify the same violation parameter in a differential measurement of our atom interferometer configuration with two different atomic species, and show that the dependence on initial conditions can be suppressed.

\subsection{UCR tests with two fountain clocks}
UCR tests can be performed by comparing two Ramsey phases $\varphi_\text{\clock}^{(j)}$, each one associated with a transition frequency $\Omega_j$ and obtained from a measurement of different species or isotopes at approximately the same location.
The differential measurement between two fountain clocks is
\begin{equation}
\label{eq:UCRFountain}
   \frac{\varphi_\text{\clock}^{(1)}}{\Omega_1T}-\frac{\varphi_\text{\clock}^{(2)}}{\Omega_2T}= \frac{\langle \vect{v}_1^2\rangle-\langle \vect{v}_2^2\rangle}{2c^2}+ \frac{\vect{g} \delta \vect{r}_0}{c^2} - \frac{\mathscr{A}}{2},
\end{equation}
where the first two contributions depend on the initial colocation difference $\delta \vect{r}_0$, as well as the difference of the second moments of velocities $\langle \vect{v}_j^2\rangle$ associated with species $j$.
These contributions limit the accuracy of UCR tests.
To avoid this issue, one has to require perfect initial colocation~\cite{Loriani2020}.
In Fig.~\ref{fig:comparison-clocks-AI}\hyperref[fig:comparison-clocks-AI]{(a)} we highlight these effects for two clocks with different initial conditions.
We observe that both interference signals $I^{(j)}_\text{\clock}$ experience a dephasing relative to each other due to imperfect colocation.

The term $\mathscr{A}$ includes UCR violations and is defined as
\begin{equation}
    \mathscr{A} = \delta \alpha \left( \frac{\vect{g} (2 \bar{\vect{r}}_0 + \bar{\vect{v}}_0T )}{c^2} - \frac{\Delta \tau}{T}\right)+ \bar \alpha \frac{\vect{g} (2 \delta \vect{r}_0 + \delta \vect{v}_0 T)}{c^2}.
\end{equation}
It depends on the difference $\delta \alpha$ of the violation parameters of both clocks and on their mean $\bar \alpha$.
Here, $\bar{\vect{r}}_0$ and $\bar{\vect{v}}_0$ are the mean initial position and velocity, respectively, and $\delta \vect{v}_0$ is the difference of initial velocities.
Apart from these means, the UCR-violation parameter $\delta \alpha$ is multiplied with $\Delta \tau$.
If UCR tests were not performed with fountain clocks but rather with trapped clocks, they would not be limited by the free-fall time or recoil effects, but by the dynamics in the trap.

In contrast to UCR tests that measure $\delta \alpha$ between two different species, tests of UGR are based on a comparison of proper times measured by clocks composed of identical atoms at different heights.
In fact, a violation is observed if this proper-time difference is modified by a factor $(1+\alpha)$.
Alternatively, such a test can be performed by comparing a freely falling clock with a stationary clock using identical atoms~\cite{Roura2020}.
Although this setup gives access to $\Delta\tau$ from Eq.~\eqref{eq:PropTime}, the measured phase is proportional to $\Omega \Delta\tau(1+\alpha)$.
Thus, despite using the same fundamental violation model, both types of experiments test different facets of local position invariance and by that of the Einstein equivalence principle.

\subsection{Atom-interferometric UCR test}
\begin{figure}
    \centering
    \includegraphics{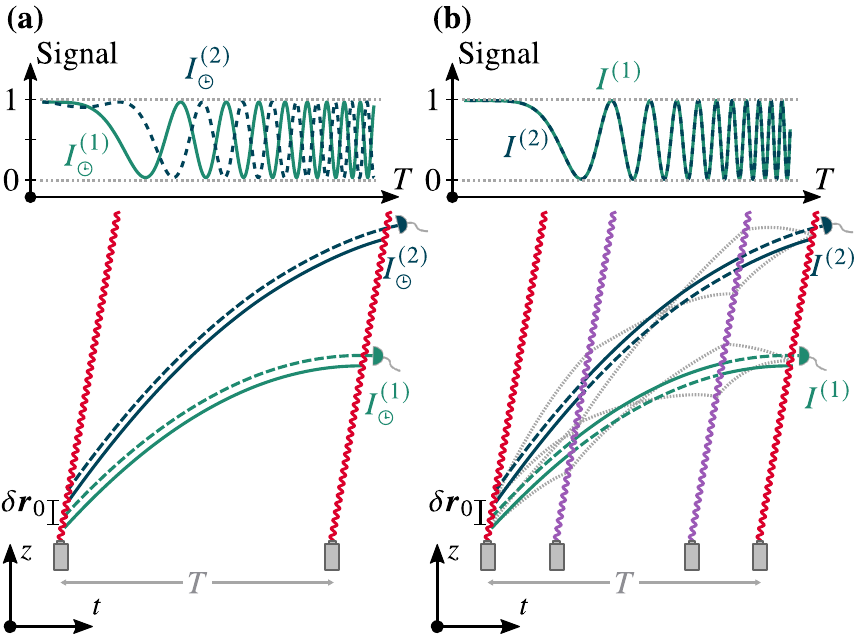}
    \caption{%
    Comparison of interference signals (top) and experimental sequences (bottom) for a fountain-clock UCR test (panel a) and the atom-interferometric UCR test (panel b).
    On the bottom of (a) we showcase the classical trajectories in a clock-based UCR test for an initial colocation mismatch of $(\delta \vect{r}_0,\delta \vect{v}_0)$ between two clocks together with the initial and final $\pi/2$ pulse of the Ramsey sequence.
    For both clocks we indicate the excited state by a dashed line and the ground state by a solid line, traveling along the same trajectory.
    The top part displays the signals $I_\text{\clock}^{(1/2)}$ corresponding to the probability to detect clock 1 or 2 in the excited state after time $T$.
    Because of the initial conditions we observe a dephasing between the signals increasing with $T$.
    In the bottom part of panel (b) we showcase the classical trajectories of the atom-interferometric UCR test together with the joint sequence of $\pi/2$-$\pi$-$\pi$-$\pi/2$ (recoilless) pulses creating two interferometers.
    We have included an initial colocation mismatch $(\delta \vect{r}_0,\delta \vect{v}_0)$ between the interferometers for both species.
    The dashed segments represent the atom in the excited state, whereas the solid ones correspond to the ground state.
    The grayed trajectories are created by pulses including a momentum recoil and are given for comparison.
    Contrary to tests with clocks and due the missing dependence on the initial conditions when using recoilless transitions, we observe no dephasing between the signals at the top of panel (b), if both species have the same transition frequency $\Omega$.
    The signals only contain a chirp due to the dependence on $\Delta \tau\propto T^3$.
    Because of the finite speed of light, this result would change for single-photon transitions.
    }
    \label{fig:comparison-clocks-AI}
\end{figure}

For our scheme shown in Fig.~\ref{fig:t3-geometry} we find from a differential measurement
\begin{equation}
\label{eq:DiffButterfly}
  \frac{\varphi^{(1)}}{\Omega_1T}-\frac{\varphi^{(2)}}{\Omega_2T}= \frac{3\Delta \tau}{16T} \frac{\delta \alpha}{2}+\frac{\bar{ \vect{v}}_r\left[ \delta  \vect{r}_0+\delta \vect{r}_T\right]}{2Tc^2}+\frac{\delta \vect{v}_r\left[ \bar{ \vect{r}}_0+ \bar{\vect{r}}_T\right]}{2Tc^2 } ,
\end{equation}
if we assume $\Omega=\omega_{\vect{k}}$ for both species and an equal propagation direction of the light beams\footnote{For recoilless transitions, the condition $\Omega=\omega_{\vect{k}}$ does not hold. 
However, Eq.~\eqref{eq:DiffButterfly} is correct, since  $\vect{v}_r=0$.}.
Here, $\bar{\vect{v}}_r$ and $\delta\vect{v}_r$ are the mean and differential recoil velocities between both species. 
Using recoilless transitions with $\vect{k}=\vect{v}_r=0$, this phase gives direct access to the UCR-violation parameter $\delta\alpha$ and solely depends on $\Delta \tau$ without initial conditions.
Thus, our proposal uses internal transitions to encode the proper-time difference $\Delta\tau$ with respect to a stationary laser in the laboratory, and compares the phases for two different atoms to read out $\Delta\tau\delta\alpha$.
We illustrate this result in Fig.~\ref{fig:comparison-clocks-AI}\hyperref[fig:comparison-clocks-AI]{(b)}.
The observed interference signals $I^{(j)}$ stay in phase for both isotopes, highlighting the robustness of our scheme against different initial conditions.

\section{Experimental considerations}
Our scheme can be performed with state-of-the-art Raman~\cite{Zhou2015,Barrett2016,Savoie2018} setups.
However, the transition frequency of a few gigahertz for rubidium~\cite{Steck2021} isotopes is not favorable, even though copropagating setups suppress parasitic recoil effects that enter in phases from finite speed of light, gravity gradients or rotations.
Contrarily, for optical $E1$-$M1$~\cite{Alden2014} or single-photon~\cite{Rudolph2020} transitions between clock states the frequency is in the terahertz range.
To discuss different experimental implementations, we focus on the case of two ytterbium~\cite{Bouganne2017} isotopes (\textsuperscript{174}Yb and \textsuperscript{176}Yb) with $\Omega= 2\pi \cdot 522$\,THz and the case of two strontium~\cite{Hu2017} isotopes (\textsuperscript{87}Sr and \textsuperscript{88}Sr) with $\Omega= 2\pi \cdot 430$\,THz.
Assuming shot-noise-limited \emph{differential} phase measurements\footnote{
For each \emph{individual} clock operated with $n$ atoms to be shot-noise limited by itself, a fractional stability of $(\Omega T \sqrt{n})^{-1}$ is necessary.
With our parameters, we arrive at $3 \cdot 10^{-20}$, which would require an improvement by about two orders of magnitude over current state of the art~\cite{Bothwell2019,Nicholson2015}.
While limiting factors arising from the trapping or the lattice are irrelevant for our proposal of freely falling clocks, the stability of interest is not the one of an individually operated clock, but the \emph{differential stability} between both clocks.
Using the same laser to address both isotopes or relying on phase locking, a differential stability between both clocks in the order of $10^{-20}$ appears to be realistic and is also required for detectors of gravitational waves or dark matter~\cite{Kolkowitz2016}.} and taking the free-fall time of soon operating 10-m fountain experiments~\cite{Schlippert2020} as a rough estimate, we find for $T= 3$\,s, $10^7$ atoms~\cite{Rudolph2020}, and $2\cdot10^6$ repetitions with an overall cycle time of $6$\,s~\cite{Schlippert2020} an uncertainty bounding the violation parameter to $\delta \alpha \leq  10^{-7}$ for ytterbium and to $\delta \alpha \leq 1.2 \cdot 10^{-7}$ for strontium in $138$-day campaigns, if colocation poses no restrictions (see below). 
This duration can be further reduced by specifically designed interleaved schemes~\cite{Savoie2018}.
For reference, current limits posed by UCR tests obtained from measurements over a 14-year period between hydrogen and cesium are at $2.2\cdot10^{-7}$~\cite{Ashby2018}, but other species have been tested as well~\cite{McGrew2019}.
A lower bound of $10^{-7}$ between ytterbium and cesium can be inferred from a combination of several violation parameters~\cite{Lange2021}, even though $\delta\alpha$ and its uncertainty has not been explicitly discussed.
Our estimates rely on established technology and benefit from the $T^3$ scaling of the phase.
Using larger atom numbers, longer measurement campaigns or squeezing techniques, this uncertainty can be improved, outperforming tests based on fountain clocks.

Besides the parasitic effects discussed below, these estimates hold for recoilless transitions with sufficiently similar isotopes.
In contrast, the finite propagation speed of the driving laser pulses cloaks UCR-violating phases for single-photon transitions, see the second term of Eq.~\eqref{eq:PhButterfly}, but drops out in a differential measurement with $\Omega=\omega_{\vect{k}}$ for both species.
For a test based on Eq.~\eqref{eq:DiffButterfly} that goes below $\delta \alpha < 10^{-7}$, initial colocation and velocity difference for single-photon transitions have to be ensured up to $\delta\vect{r}_0<4.2\,\text{mm}$ and $\delta\vect{v}_0< 1.4\,\text{mm}/\text{s}$ for ytterbium and up to $\delta\vect{r}_0<2.5\,\text{mm}$ and $\delta\vect{v}_0< 0.8\,\text{mm}/\text{s}$ for strontium, which are not ambitious requirements.
Given that $\delta \vect{v}_r$ is two orders of magnitude smaller than $\bar{\vect{v}}_r$, no additional restrictions arise from the last term of Eq.~\eqref{eq:DiffButterfly} if the mean initial position and velocity are known to the precision stated above.
Even though recoilless transitions are favorable, single-photon transitions do not intrinsically lead to ambitious requirements, so that the scheme is robust against initial conditions.
Besides, no further UCR violations arise from light propagation in gravity in a UCR-violating background~\cite{DiPumpo2022}. 

Additional parasitic perturbations may involve different accelerations for both internal states for example caused by coupling to electromagnetic fields~\cite{Amit2019}, i.e., the Zeeman and/or Stark effect.
Other contributions are given by gravity gradients~\cite{Roura2014,Roura2017,Overstreet2018,Ufrecht2021,Zimmermann2021} or other harmonic potentials, which give rise to relevant phases for a significant momentum transfer.
Both state-dependent accelerations and harmonic potentials cause different accelerations on the two branches of the interferometer and, therefore, cloak UCR violations.
Additional deleterious effects are introduced by rotations~\cite{Lan2012,Dickerson2013}.
See the \hyperref[app:a]{Appendix} for a derivation of these phase contributions.

As a consequence of the linearity of first-order perturbation theory, $\delta \alpha$ in Eq.~\eqref{eq:DiffButterfly} is replaced by $\delta \alpha+\epsilon$ for differential UCR tests, where the perturbations are included in a dimensionless parameter $\epsilon$.
We observe that for both ytterbium and strontium no contributions from rotations occur, assuming that a nonperturbative, leading-order influence of rotations is already compensated~\cite{Lan2012,Dickerson2013} by adjusting the wave vector in case of single-photon transitions or via tip-tilt mirror systems for $E1$-$M1$ or counterpropagating Raman transitions.
Even though residual rotations do not contribute in this differential scheme, it is necessary to mitigate the dominant contribution to justify our perturbative treatment.

Residual accelerations $\delta\vect{a}$ between isotopes with sufficiently similar polarizability and black-body-radiation shifts (like \textsuperscript{174}Yb and \textsuperscript{176}Yb or \textsuperscript{87}Sr and \textsuperscript{88}Sr considered in our estimates) result in a nonvanishing effect $\vect{g}\delta \vect{a}/\vect{g}^2$.
For such state-dependent forces the projection $\vect{g}\delta\vect{a}$ has to be limited at the same level as the required precision of $\vect{g}^2\delta\alpha$.
This condition gives rise to $|\delta \vect{a}|<10^{-6}\text{m}/\text{s}^2$ both for ytterbium and strontium for $\epsilon < 10^{-7}$.
In fact, these residual accelerations pose a limit for both recoilless and single-photon transitions.

A quadratic potential in turn leads to an additional dependence on initial conditions, setting more demanding requirements on the initial velocity difference $\delta\vect{v}_0$ between the isotopes under consideration.
While no spatial separation is generated from recoilless transitions with $\vect{k}=0$ and consequently no phase difference is introduced by quadratic potentials, we find the ambitious constraint $|\delta \vect{v}_{0}|<10^{-8}\text{m}/\text{s}$ both for ytterbium and strontium for a limit of $\epsilon<10^{-7}$ in Earth-based setups when relying on single-photon transitions.
However, one can apply mitigation schemes~\cite{Roura2014,Roura2017,Overstreet2018,Ufrecht2021,Zimmermann2021} to further suppress gravity gradients.
In this case, the residual gradient is not the original gravity gradient of Earth but has to be replaced by its uncertainty remaining after compensation and is further suppressed~\cite{Loriani2020} by several orders of magnitude.
Gravity gradients therefore pose the main limitation to measurements performed with single-photon transitions, also compared to finite-speed-of-light effects, while they are irrelevant for $E1$-$M1$ transitions.

Isotopes or species with considerably differing transition frequencies lead to generalized expressions and can be easily obtained based on our analytical treatment (see the \hyperref[app:a]{Appendix}).
However, no additional terms arise for phase contributions that scale linearly with the transition frequency and that are independent of the recoil, due to the form of the differential measurement from Eq.~\eqref{eq:DiffButterfly}.
In contrast, other contributions are suppressed by the ratio of transition frequency difference and mean transition frequency.
Additional phase fluctuations, e.g., caused by mirror vibrations in retroreflective setups and $E1$-$M1$ transitions or by laser phase jitter, contribute but are suppressed by this factor.
Working with sufficiently similar species or isotopes circumvents these issues entirely.

\section{Comparison and contextualization}
Some of these parasitic effects also affect UCR tests with fountain clocks and give rise to similar modifications as discussed for our interferometer geometry.
Moreover, Eq.~\eqref{eq:UCRFountain} does not include recoil effects which dominate for single-photon transitions and for long time intervals $T$, leading to displaced wave packets and diminishing contrast~\cite{Roura2014} when applied to clocks.
Usual UCR tests in fountain clocks are not performed with such optical transitions in free fall, since they would require traps to suppress recoil effects in the Lamb-Dicke regime.
Consequently, such tests are only accessible in the gigahertz range of the internal transition.
In contrast to deleterious effects in atom interferometers, the contributions arising from initial conditions and second moments in Eq.~\eqref{eq:UCRFountain} cannot be straightforwardly compensated even for recoilless transitions.

Violations of the Einstein equivalence principle are also included in Mach-Zehnder~\cite{Giulini2012,DiPumpo2021} setups performed with Raman diffraction or single-photon transitions.
In analogy to above such effects are suppressed by the small transition frequency in Raman-based schemes.
Moreover, any Mach-Zehnder geometry with transitions between internal states at each pulse is open at the end of the sequence~\cite{Roura2020,DiPumpo2021}.
Consequently, initial conditions contribute even without the deleterious effects and finite speed of light studied in our article.
Besides, the dominating~\cite{Kasevich1991} contribution $-\vect{k}\vect{g}T^2$ in the Mach-Zehnder configuration makes it challenging to isolate the desired phase.
Nevertheless, this configuration can be used for tests of different aspects of the equivalence principle~\cite{DiPumpo2021}.

Tests of UGR with atom interferometers are limited by the dimensions of the experiment so that (trapped) atomic clocks have an inherent advantage~\cite{DiPumpo2021}.
Our proposed test of LPI based on atom interferometry has a sensitivity competitive to atomic fountain clocks and the first one with the potential to outperform a classical, localized EEP test.
The technology enabling such experiments, possibly combined with large-momentum-transfer techniques, can be also used for tests of the universality of free fall, so that such an experimental facility may be fit to test two facets of the Einstein equivalence principle. 

\section*{Acknowledgments}
We are grateful to W. P. Schleich for his stimulating input and continuing support.
We also thank S. Abend, A. Bott, M. A. Efremov, N. Huntemann, R. Lopp, E. M. Rasel, D. Schlippert, C. Schubert, W. G. Unruh, A. Wolf, M. Zimmermann as well as the QUANTUS, INTENTAS, and VLBAI teams for fruitful and interesting discussions.
The projects ``Metrology with interfering Unruh-DeWitt detectors'' (MIUnD) and ``Building composite particles from quantum field theory on dilaton gravity'' (BOnD) are funded by the Carl Zeiss Foundation (Carl-Zeiss-Stiftung).
The QUANTUS and INTENTAS projects are supported by the German Space Agency at the German Aerospace Center (Deutsche Raumfahrtagentur im Deutschen Zentrum f\"ur Luft- und Raumfahrt, DLR) with funds provided by the Federal Ministry for Economic Affairs and Climate Action (Bundesministerium f\"ur Wirtschaft und Klimaschutz, BMWK) due to an enactment of the German Bundestag under Grant Nos. 50WM1956 (QUANTUS V), 50WM2250D-2250E (QUANTUS+), as well as 50WM2177-2178 (INTENTAS).
E.G. thanks the German Research Foundation (Deutsche Forschungsgemeinschaft, DFG) for a Mercator Fellowship within CRC 1227 (DQ-mat).

\appendix
\section*{Appendix}
\subsection{Formalism to describe atom interferometry}
\label{app:a}
A two-level atom consists of a ground state $\ket{g}$ and an excited state $\ket{e}$, fulfilling the completeness relation $\mathds{1}_\text{int}=\ket{e}\!\bra{e} +\ket{g}\!\bra{g}$.
Driving transitions between these internal states, for example by the interaction with a laser, may give rise to a momentum transfer $\pm\hbar\vect{k}$ based on energy-momentum conservation~\cite{Giese2015}.
This way, one generates beam splitters and mirrors for matter waves. 
The interaction with pulse $\ell$ is effectively described~\cite{Kleinert2015,Kleinert2019} by
\begin{subequations}
\begin{align}
    \hat{U}_\text{BS}^{(\ell)}=& \frac{1}{\sqrt{2}}\left(\mathds{1}_\text{int}  - \ii\ee^{\ii\vect{k}_\ell\hat{\vect{r}}} \ket{e}\!\bra{g}- \ii  \ee^{-\ii\vect{k}_\ell\hat{\vect{r}}} \ket{g}\!\bra{e}\right) \\
    \hat{U}_\text{M}^{(\ell)}= & - \ii\ee^{\ii\vect{k}_\ell\hat{\vect{r}}} \ket{e}\!\bra{g}- \ii  \ee^{-\ii\vect{k}_\ell\hat{\vect{r}}} \ket{g}\!\bra{e},
\end{align}
\end{subequations}
which transfers a momentum $\pm \hbar \vect{k}_\ell$ due to the photonic recoil upon an internal transition.
Here, $\hat{\vect{r}}$ is the position operator and we have omitted effects of laser phases and finite pulse durations.
The operator $\hat{U}_\text{BS}^{(\ell)}$ describes a $\pi/2$ pulse, while $\hat{U}_\text{M}^{(\ell)}$ acts as a $\pi$ pulse.

An arbitrary interferometer sequence $\hat{U}_\text{seq}$ is given by combinations of $\hat{U}_\text{BS}^{(\ell)}$ and $\hat{U}_\text{M}^{(\ell)}$ and the evolution of the atom in external potentials between those pulses.
The latter is diagonal in the internal degrees of freedom for long-lived states.
Hence, the pulse sequence $\hat{U}_\text{seq}$ is acting on the internal states as well as the center-of-mass (c.m.) motion.
For the setups discussed in this article, the relevant observable is given by a projection $\hat\Pi=\ket{j}\!\bra{j}$ on the internal states $j=e,g$, leading to the  interference signal~\cite{Kleinert2015,Kleinert2019,DiPumpo2021}
\begin{equation}
    I = \text{Tr}_\text{c.m.}\text{Tr}_\text{int}\left(\hat{U}^\dagger_\text{seq} \hat{\Pi}\hat{U}_\text{seq}\rho_\text{int}(0)\otimes \rho_\text{c.m.}(0)\right).
\end{equation}
Here, we assumed that the initial state of the internal degrees of freedom $\rho_\text{int}(0)$ and the initial state of c.m. motion $\rho_\text{c.m.}(0)$ are uncorrelated.
We further assume that all atoms are initially in the ground state $\rho_\text{int}(0)=\ket{g}\!\bra{g}$ and that a measurement of the excited-state population is performed at the end of the sequence, hence we identify $\hat\Pi=\ket{e}\!\bra{e}$.
Carrying out the partial trace
\begin{equation}
    \text{Tr}_\text{int}\left[\hat{U}^\dagger_\text{seq} \hat{\Pi}\hat{U}_\text{seq}\rho_\text{int}(0)\right]= \bra{e}\hat{U}_\text{seq}\ket{g}^\dagger \bra{e}\hat{U}_\text{seq}\ket{g},
\end{equation}
we find that the matrix element $\bra{e}\hat{U}_\text{seq}\ket{g}=(\hat{U}_1+\hat{U}_2)/2$ consists of a superposition of two branches $\sigma=1,2$ acting on the c.m. motion.
Each branch can be assigned to a unitary time evolution $\hat{U}_\sigma^\dagger \hat{U}_\sigma =\mathds{1}_\text{c.m.}$, associated with an effective Hamiltonian $\hat{H}^{(\sigma)}$ for a branch-dependent description.
With this insight, we directly obtain the interference signal
\begin{equation}
    I =  \frac{1}{4}\left[2 + \text{Tr}_\text{c.m.}\left(\hat{U}^{\dagger}_1\hat{U}^{}_2 \rho_\text{c.m.}(0)+\text{h.c.}\right)\right]=  \frac{1}{2} (1 + C \cos \varphi)
\end{equation}
with the expectation value $\big<\hat{U}^{\dagger}_1\hat{U}^{}_2\big>=C \exp({\ii\varphi})$ of the overlap operator $\hat{U}^{\dagger}_1\hat{U}^{}_2$.
Here, we defined the contrast $C$ and phase $\varphi$.

The effective, branch-dependent Hamiltonian $\hat{H}^{(\sigma)} = \hat{H}^{(\sigma)}_0 + \hat{\mathcal{H}}^{(\sigma)}$ can be divided into the dominant contribution
\begin{equation}
    \hat{H}^{(\sigma)}_0=m c^2+\frac{\hat{\vect{p}}^2}{2m}+m \vect{g}\hat{\vect{r}}-\hbar\sum_\ell{\left(\vect{k}^{(\sigma)}_\ell\hat{\vect{r}}-\omega^{(\sigma)}_\ell t\right)\delta(t-t_\ell)}
\end{equation}
and a term $\hat{\mathcal{H}}^{(\sigma)}$ that describes state-dependent effects and other perturbations.
Here, $m$ is the atom's mean mass between ground and excited state, $\vect{g}$ the gravitational acceleration, $\hat{\vect{p}}=\left(\hat{p}_x,\hat{p}_y,\hat{p}_z\right)^{\text{T}}$ the momentum operator, and $\vect{\hat{r}}=\left(\hat{x},\hat{y},\hat{z}\right)^{\text{T}}$ the position operator, fulfilling $[\hat{r}_i,\hat{p}_j]=\ii \hbar \delta_{ij}$ with Kronecker delta $\delta_{ij}$.
Moreover, $\hbar \vect{k}_\ell^{(\sigma)}$ describes the momentum transfer induced by the $\ell$th pulse at time $t_\ell$ on branch $\sigma$, associated with the (effective) wave vector of the diffracting light field.
Similarly, $\omega_\ell^{(\sigma)}$ denotes the (effective) frequency of the pulse, where its sign depends on the pulse and branch and can be derived from the expression of the laser phase.
\begin{figure*}[t]
    \centering
    \includegraphics{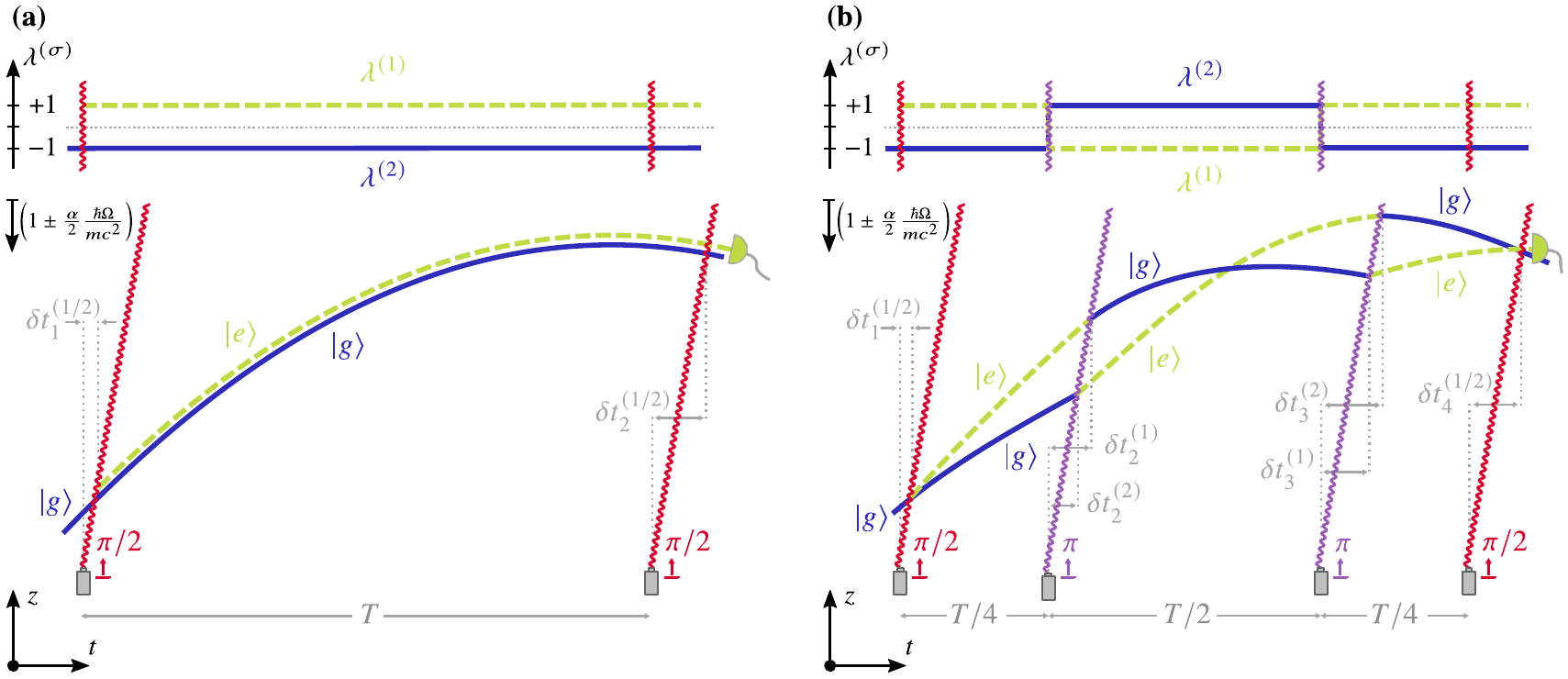}
    \caption{
    State-dependent functions $\lambda^{(\sigma)}(t)$ (top) and trajectories in spacetime diagrams (bottom) for the calculation of the phase shifts.
    Panel (a) shows an atomic clock in a Ramsey sequence, where a $\pi/2$ pulse (red) initiates a superposition of internal states (ground state in blue and excited state in green) and another $\pi/2$ pulse after a time $T$ that interferes them.
    The finite propagation speed of both light pulses is highlighted by inclined lines and leads to time delays $\delta t_\ell^{(1/2)}$ common for both internal states.
    During the sequence, the (unperturbed) atom remains in a superposition of both internal states so that both $\lambda^{(\sigma)}(t)$ are constant.
    Since there is no momentum transfer, both wave packets corresponding to the two internal states are centered along the same spacetime trajectory.
    Panel (b) shows the atom interferometer, where a $\pi/2$ pulse (red) initiates a superposition of both internal states entangled with corresponding c.m. states by transferring a momentum $\hbar \vect{k}$ in addition to the internal transition.
    Two $\pi$ pulses (purple) at times $T/4$ and $3T/4$ invert momenta and internal states, before a last $\pi/2$ pulse at time $T$ interferes both branches.
    We showcase the finite speed of light by inclined lines for the pulses, which leads to time delays $\delta t_\ell^{(\sigma)}$ that differ for both branches.
    The top shows the (unperturbed) sequence of internal states of both branches encoded in the functions $\lambda^{(\sigma)}(t)$ that jump by $\pm 2$ during the $\pi$ pulses.
    In contrast to atomic clocks, both internal states travel along different branches.
    }
    \label{fig:app:t3-vs-clock}
\end{figure*}

\subsection{Perturbations}
\label{app:b}
The perturbative term
\begin{align}
\begin{split}
    \label{eq:PerturPot}
    \hat{\mathcal{H}}^{(\sigma)}=& \lambda^{(\sigma)}(t)\frac{\hbar\Omega}{2 c^2}\left[c^2-\frac{\hat{\vect{p}}^2}{2 m^2}+\left(1+\alpha\right)\vect{g}\hat{\vect{r}}\right]+\hat{V}^{(\sigma)}_\text{FL}+\hat{V}^{(\sigma)}_\text{del}
\end{split}
\end{align}
includes a coupling of the internal degrees of freedom to the c.m. motion with a factor $\hbar \Omega /c^2$, which represents a consistent treatment of relativistic corrections due to the mass-energy relation~\cite{Zych2011,Sonnleitner2018,Schwartz2019}.
The parametrization of violations of the Einstein equivalence principle is given by a modified gravitational acceleration $\pm \left(1+\alpha\right)\vect{g}$ for different internal states and can be derived for example from dilaton models~\cite{Alves2000,Damour1999,Damour2010,Damour2012,Roura2020,Ufrecht2020,DiPumpo2021,DiPumpo2022}.
This modification implies violations of the Einstein equivalence principle in its different facets:
(i) Assuming energy conservation, this factor can be connected to violations of UFF~\cite{Will2014,DiPumpo2021}.
However, in such a setting it constitutes a test that compares the gravitational acceleration of atoms in different internal states, rather than two different species or test objects.
(ii) Inferring the factor $(1+\alpha)$ from two clocks composed of the same species at different heights leads to tests of UGR.
(iii) Measuring a difference $\delta\alpha$ of two parameters $\alpha$, each one associated with a different species positioned at the same height, gives rise to the UCR tests at the heart of our proposal.
Although the factor $\alpha$ itself appears in all of these tests based on different internal states, it is the exact form and combination in the phase of interest which defines what kind of test is performed.
Moreover, the time- and branch-dependent function $\lambda^{(\sigma)}(t)$ is given by $\lambda^{(\sigma)}(t)=1$ while the (unperturbed) atom propagates in the excited state and by $\lambda^{(\sigma)}(t)=-1$ when propagating in the ground state.

Effects of the finite propagation velocity of laser pulses~\cite{Dimopoulos2008,Tan2016,Tan2017} are encoded in
\begin{equation}
    \hat{V}_\text{FL}=-\hbar \sum_{\ell}{\left(\vect{k}_\ell\hat{\vect{r}}-\omega_\ell t\right)\left[\delta(t-t_\ell-\delta t_\ell)-\delta(t-t_\ell)\right]},
\end{equation}
where we suppressed the superscripts $\sigma$ of $\vect{k}$, $\omega_\ell$ and $\delta t_\ell$.
Here, $\delta t_\ell^{(\sigma)} $is the time delay due to the finite propagation time of light from the origin of the laser to the branch, see Fig.~\ref{fig:app:t3-vs-clock}, and is treated as a perturbative quantity.
Additional branch-dependent phase contributions $ \pm \Omega  \delta t_\ell^{(\sigma)} /2 $ arise from the leading term of Eq.~\eqref{eq:PerturPot} including finite speed of light.

We include further deleterious effects that act as a perturbation in $\hat{V}_\text{del}^{(\sigma)}$, which is given by
\begin{align}
\begin{split}
    \hat{V}^{(\sigma)}_\text{del}=\lambda^{(\sigma)}(t)\dfrac{\hbar \Omega}{2} \dfrac{\vect{a}\hat{\vect{r}}}{c^2}+\dfrac{m}{2}\hat{\vect{r}}^{\text{T}}\Gamma\hat{\vect{r}}-\vect{\omega}_\text{rot}\left(\vect{\hat{r}}\times\hat{\vect{p}}\right),
\end{split}
\end{align}
where $\vect{a}$ is a differential acceleration between both internal states, for example caused by Zeeman or Stark effects.
Here, $\Gamma$ is a matrix containing squares of frequencies of a harmonic potential, which could be caused by (partially compensated) gravity gradients~\cite{Roura2014,Roura2017,Overstreet2018,Ufrecht2021,Zimmermann2021}.
Moreover, $\vect{\omega}_\text{rot}$ is the rotation frequency of the laboratory, where nonperturbative contributions are assumed to be already mitigated~\cite{Lan2012,Dickerson2013}.
Effects from relativistic corrections to the c.m. motion, such as terms including $\hat{\vect{p}}^4/c^2$ and $\left(\vect{g}\hat{\vect{z}}\right)\hat{\vect{p}}^2/c^2$, lead to contributions which are further suppressed and thus can be neglected.
Similar arguments also apply for the influence of modified wave vectors~\cite{DiPumpo2022} in case of the schemes considered in this article.
\begin{table*}[htb]
\caption{Perturbative contributions to the proposed atom interferometer scheme.
The left column summarizes the contributions to the classical counterpart $\mathcal{H}^{(\sigma)}$ of the perturbative Hamiltonian. 
This Hamiltonian includes constant, kinetic, and gravitational couplings to the transition frequency $\Omega$, as well as state-dependent differential accelerations $\vect{a}$, finite-speed-of-light effects that introduce a time delay $\delta t^{(\sigma)}_\ell$, an additional harmonic oscillator with frequency matrix $\Gamma$, and rotations of frequency $\vect{\omega}_\text{rot}$.
The perturbative derivation of the individual phase contributions includes position trajectories $\vect{r}^{(\sigma)}(t)$ and their derivatives $\dot{\vect{r}}^{(\sigma)}(t)$.
The right column shows the corresponding effects on the phase shift $\varphi_1$.
Whereas the term which couples solely to $\Omega$ yields no contribution, the other two terms involving such a coupling basically lead to the UCR violating phase of Eq.~\eqref{eq:PhButterfly}.
The term including the state-dependent differential acceleration yields a similar contribution scaling with $T^3$, as it is also the case for rotations.
In contrast, the effects from the harmonic potential include a cubic as well as a quartic scaling with $T$.
Finite speed of light yields contributions including the frequency $\omega_{\vect{k}}=c\left|\vect{k}\right|$, the gravitational acceleration $g_{\vect{k}}=\vect{k}\vect{g}/\left|\vect{k}\right|$ in the direction of light propagation, and the mean trajectory $\vect{r}_T=\vect{r}_0+\vect{v}_0 T+\vect{v}_r T/2-\vect{g}T^2/2$ at time $T$.
Here, we defined the recoil velocity vector $\vect{v}_r=\hbar\vect{k}/m$.
The result is valid to linear order in the time delay $\delta t^{(\sigma)}_\ell =\vect{k}\vect{r}^{(\sigma)}(t_\ell)/ (c|\vect{k}|)$.
}
\label{tab:exp2}
\begin{center}
\begin{tabular}{llr}
\toprule
Physical effect  &$\mathcal{H}^{(\sigma)}$ & $\varphi_1$  \\ 
\midrule
Internal energy &$\lambda^{(\sigma)}\dfrac{\hbar\Omega}{2}$ & $0$ \\[2ex]
Kinetic mass-energy coupling &$-\lambda^{(\sigma)}\dfrac{\hbar\Omega}{2 c^2}\dfrac{m ( \dot{\vect{r}}^{(\sigma)})^2}{2}$ & $\dfrac{\Omega T^3\vect{g}^2}{32c^2}$ \\[2ex]
Gravitational mass-energy coupling &$\lambda^{(\sigma)}\dfrac{\hbar\Omega}{2 c^2}\left(1+\alpha\right)\vect{g}\vect{r}^{(\sigma)}$ & $(1+\alpha)\dfrac{\Omega T^3\vect{g}^2 }{32c^2}$ \\[2ex]
Finite speed of light &$-\hbar \displaystyle\sum\limits_\ell \delta t_\ell^{(\sigma)} \left(\vect{k}_\ell^{(\sigma)}\dot{\vect{r}}^{(\sigma)}-\omega_\ell^{(\sigma)} \right) \delta(t-t_\ell) $ & $-\frac{3\omega_{\vect{k}} T^3g^2_{\vect{k}}}{32c^2}+\frac{\omega_{\vect{k}}\vect{v}_r\left[\vect{r}_0+\vect{r}_T\right]}{2c^2} $ \\[2ex]
State-dependent accelerations &$\lambda^{(\sigma)} \dfrac{\hbar \Omega}{2 c^2} \vect{a}\vect{r}^{(\sigma)}$ & $\dfrac{\Omega T^3\vect{g} \vect{a}}{32 c^2}$ \\[2ex]
Gravity gradients &$\dfrac{m}{2}\vect{r}^{(\sigma)\text{T}}\Gamma\vect{r}^{(\sigma)}$& $\dfrac{1}{32}\vect{k}^{\text{T}} \Gamma \left(\vect{v}_0+\frac{\vect{v}_r}{2}-\dfrac{1}{2}\vect{g}T\right)T^3$  \\[2ex]
Rotations &$-m\vect{\omega}_\text{rot}\left(\vect{r}^{(\sigma)}\times\dot{\vect{r}}^{(\sigma)}\right)$ &$\dfrac{1 }{16}\vect{g}\left(\vect{\omega}_\text{rot}\times\vect{k}\right) T^3$ \\
\bottomrule
\end{tabular}
\end{center}
\end{table*}

We derive the phase $\varphi=\varphi_0+\varphi_1+\varphi_\text{WP}$ from perturbative methods~\cite{Ufrecht2019,Ufrecht20202}, where $\varphi_0$ is the unperturbed phase shift generated by $\hat{H}_0^{(\sigma)}$ and can be obtained from trajectories $\vect{r}^{(\sigma)}(t)$ generated by the classical counterpart of $\hat{H}_0^{(\sigma)}$.
Inserting these classical trajectories into the perturbation $\hat{\mathcal{H}}^{(\sigma)}$ leads to the phase 
\begin{equation}
    \label{eq:GeneralPhasePerturb}
    \varphi_1=-\frac{1}{\hbar}\int\limits_0^T\!\text{d}t\,\left(\mathcal{H}^{(1)}-\mathcal{H}^{(2)}\right).
\end{equation}
Wave-packet effects $\varphi_\text{WP}$ are generated by different deformations of wave packets due to perturbations, leading to a nonperfect overlap at the end of the interferometer.
Generalizing the technique from Refs.~\cite{Ufrecht2019,Ufrecht20202} to perturbations consisting of kinetic terms and rotations, we find
\begin{align}
\label{PerturbativeWPEffects}
\begin{split}
    \varphi_\text{WP}=-\frac{1}{2\hbar}\oint\!\text{d}t\,&\Big\lbrace\partial_r^2\mathcal{H}^{(\sigma)}_{ij}\left<\hat{r}^i_c\hat{r}^j_c\right>+\partial^2_p\mathcal{H}^{(\sigma)}_{ij}\left<\hat{p}^i_c \hat{p}^j_c\right>\\
    &+\partial_r\partial_p\mathcal{H}^{(\sigma)}_{ij}\left<\left\lbrace\hat{p}^i_c, \hat{r}^j_c\right\rbrace\right> \Big\rbrace
\end{split}
\end{align}
where the line integration goes along branch (1) from initial time to final time and then back in time along path (2).
The centered time-dependent operators $\hat{\vect{r}}_c(t)=\hat{\vect{r}}-\vect{r}_0+(\hat{\vect{p}}/m-\vect{v}_0)t$ and $\hat{\vect{p}}_c=\hat{\vect{p}}-m\vect{v}_0$ have vanishing expectation values.
The derivatives in Eq.~\eqref{PerturbativeWPEffects} are defined by $\partial^2_p\mathcal{H}^{(\sigma)}_{ij}=-\delta_{ij}\lambda^{(\sigma)}(t)\hbar\Omega/(2m^2 c^2)$, $\partial_r\partial_p\mathcal{H}^{(\sigma)}_{ij}=-\omega^m_\text{rot}\epsilon_{mij}$ with $\epsilon_{mij}$  being the Levi-Civita tensor, and $\partial^2_r\mathcal{H}^{(\sigma)}_{ij}=m\Gamma_{ij}$.
The latter two do not contribute to the signal of clocks and atom interferometers because $-\omega^m_\text{rot}\epsilon_{mij}$ and $m\Gamma_{ij}$ as well as $\hat{\vect{r}}_c(t)$ and $\hat{\vect{p}}_c$ do not depend on the internal state or branch, and cancel upon integration.
Hence, only wave-packet contributions from the derivative with respect to momenta arise. 

\subsection{Atomic clocks}
\label{app:c}
For clocks, we choose $\vect{k}_\ell^{(\sigma)}=0$ (neglecting recoils) for all pulses, as well as $\lambda^{(1)}(t)=1$ and $\lambda^{(2)}(t)=-1$ for the whole duration of a Ramsey sequence, see Fig.~\ref{fig:app:t3-vs-clock}\hyperref[fig:app:t3-vs-clock]{(a)} on the top.

The figure shows on the bottom the unperturbed classical trajectory through space for freely falling atoms.
It is given by $\vect{r}(t)=\vect{r}_0+\vect{v}_0t-\vect{g}t^2/2$ and the velocity $\vect{v}(t)=\vect{v}_0-\vect{g}t$ for both internal states with initial position $\vect{r}_0$ and initial velocity $\vect{v}_0$.
We find $\varphi_0=0$ except for laser phases that may arise from chirping or vibrations and hence omit this term.

Because $\lambda^{(1)}(t)-\lambda^{(2)}(t)=2$ for all times, we find on resonance
\begin{align}
\begin{split}
    \varphi_1=- \frac{\Omega}{ c^2} \int\limits_0^T \dd t \left[-\frac{\vect{v}^2(t)}{2}+\left(1+\alpha\right)\vect{g}\vect{r}(t)\right]
\end{split}
\end{align}
if we neglect the influence of any further deleterious effects for clocks.
In principle, those contributions can be incorporated.
We observe that $\varphi_1$ corresponds exactly to $-\Omega (\tau-T)$, if the proper-time difference from Eq.~\eqref{eq:PropTime} is modified by a parameter $\alpha$ and taking nonvanishing initial conditions into account.
The integration along the classical trajectories gives rise to
\begin{align}
\begin{split}
    \frac{\varphi_1}{-\Omega T}=&\left(1+\frac{\alpha}{2}\right)\left[\frac{\vect{g}(2\vect{r}_0+\vect{v}_0 T)}{c^2}-\frac{\Delta \tau}{T}\right] - \frac{\vect{g}\vect{r}_0}{ c^2}-\frac{\vect{v}_0^2}{2c^2}.
\end{split}
\end{align}
With the same method, we find for the wave-packet effects
\begin{equation}
    \varphi_\text{WP}=\frac{\Omega}{2 c^2}\int\limits_0^T\!\text{d}t\,\frac{\Delta \vect{p}_0^2}{m^2}=\frac{\Omega T}{2 c^2}\left(\langle \vect{v}_0^2\rangle-\vect{v}_0^2\right),
\end{equation}
from which we obtain the result presented in Eq.~\eqref{eq:phase_clock} by adding $\phi_1$ and $\phi_\text{WP}$.
\vspace{-1em}
\subsection{Atom interferometer}
\label{app:c}
The atom interferometer introduced in the article corresponds to a butterfly or figure-eight geometry~\cite{Clauser1988,Marzlin1996,McGuirk2002,Kleinert2015}.
Figure~\ref{fig:app:t3-vs-clock}\hyperref[fig:app:t3-vs-clock]{(b)} shows on the bottom a spacetime diagram of this scheme and both branch-dependent trajectories~\cite{DiPumpo2021}
\begin{align}
\begin{split}
    \vect{r}^{(\sigma)}(t)=&\,\vect{r}^{(\sigma)}(t_\ell)+\left[\dot{\vect{r}}^{(\sigma)}(t_\ell)+\frac{\hbar}{m} \vect{k}^{(\sigma)}_\ell\right]\left(t-t_\ell\right)\\
    &-\frac{1}{2}\vect{g}\left(t^2-t^2_\ell\right)
\end{split}
\end{align}
valid in the segment $t_{\ell+1}>t\geq t_\ell$.
This expression is more involved compared to clocks due to the momentum recoil that generates a spatial superposition.
Both branches have the same initial conditions right before the first pulse, i.e., $\vect{r}^{(1)}(0)=\vect{r}^{(2)}(0)=\vect{r}_0$ and $\dot{\vect{r}}^{(1)}(0)=\dot{\vect{r}}^{(2)}(0)=\vect{v}_0$.
From the figure, we identify the momentum transfer of the individual pulses $\vect{k}= \vect{k}_1^{(1)}= -\vect{k}_2^{(1)}=\vect{k}_3^{(1)}=\vect{k}_2^{(2)}=-\vect{k}_3^{(2)} = \vect{k}_4^{(2)} $ of the sequence, where the laser frequencies $\omega^{(\sigma)}_\ell$ are defined analogously.
Due to the symmetry of these transfers, and by that due to the trajectories, no contributions to $\varphi_0$ arise.
The functions $\lambda^{(\sigma)}(t)$ that describe the two internal states and their transitions are plotted in Fig.~\ref{fig:app:t3-vs-clock}\hyperref[fig:app:t3-vs-clock]{(b)} on the top.
The time delays $\delta t_\ell^{(\sigma)}=\vect{k}\vect{r}^{(\sigma)}(t_\ell)/ (c|\vect{k}|)$ that arise from the finite speed of light propagation are also shown in the figure. 

Table~\ref{tab:exp2} summarizes the results of the integration of each of the individual terms that contribute to $\varphi_1$ and give rise to the equations used in the main body of the article.
The results for finite speed of light have been obtained from an expansion to linear order in $\delta t_\ell^{(\sigma)}.$
No contributions from wave-packet effects are contributing because the atom remains in each internal state for the same time on each branch, which can also be seen from $\int_0^T\dd t\lambda^{(\sigma)}(t) =0$, as apparent from Fig.~\ref{fig:app:t3-vs-clock}\hyperref[fig:app:t3-vs-clock]{(b)}.

\bibliography{Literatur}

\end{document}